\documentclass[twocolumn,aps,preprintnumbers,amsmath,amssymb]{revtex4}

\usepackage{graphicx}
\usepackage{dcolumn}
\usepackage{bm}

\usepackage{amsmath}
\usepackage{graphicx}
\usepackage{amssymb}
\usepackage{subfig}
\usepackage[utf8]{inputenc}
\usepackage{comment}
\usepackage{verbdef}
\verbdef{\vtext}{p p > Z' j}
\verbdef{\vtexta}{p p > Z' a}
\verbdef{\vtextb}{p p > Z'}
\usepackage{filecontents}

\begin{document}


\title{A Bottom Line for the LHC Data by Leveraging Pileup as a Zero Bias Sample}

\author{Benjamin Nachman}
 \email{bnachman@cern.ch}
\affiliation{Lawrence Berkeley National Laboratory}

\author{Francesco Rubbo}
 \email{rubbo@cern.ch}
\affiliation{SLAC National Accelerator Laboratory}

\date{\today}

\begin{abstract}
Due to a limited bandwidth and a large proton-proton interaction cross-section relative to the rate of interesting physics processes, most events produced at the Large Hadron Collider (LHC) are discarded in real time.  A sophisticated trigger system must quickly decide which events should be kept and is very efficient for a broad range of processes.  However, there are many processes that cannot be accommodated by this trigger system.  Furthermore, there may be models of physics beyond the Standard Model (BSM) constructed after data taking that could have been triggered, but no trigger was implemented at run time.  Both of these cases can be covered by exploiting pileup interactions as an effective zero bias sample.   At the end of High-Luminosity LHC operations, this zero bias dataset will have accumulated about 1 fb$^{-1}$ of data from which a bottom line cross-section limit of $\mathcal{O}(1)$ fb can be set for BSM models already in the literature and those yet to come.

\end{abstract}

\maketitle

\section{Introduction}
\label{sec:intro}

At a proton-proton ($pp$) collider like the Large Hadron Collider (LHC), interesting events are rare.
Unlike electron-positron colliders, the partonic center-of-mass energy $\sqrt{\hat{s}}$
follows a broad distribution set by parton distribution functions (PDF).
As such, the total inelastic cross-section ($\mathcal{O}(100)$ mb~\cite{Aaboud:2016mmw,CMS:2016ael} at $\sqrt{s}=13$ TeV)
is many orders of magnitude above the production cross-section for electroweak scale particles,
such as the $W$ boson ($\mathcal{O}(10)$ nb~\cite{Aad:2016naf,CMS:2015ois} at $\sqrt{s}=13$ TeV).
In order to increase the rate of interesting events as much as possible, the LHC is operated at very high luminosities.
Proton bunches collide every 25 ns and the bunch density is such that multiple $pp$ interactions (pileup) occur in each bunch crossing.
Due to limited readout and disk space capabilities, it is not possible to fully record every bunch collision.
Therefore, both the ATLAS and CMS experiments have developed strategies to trigger on events of interest.
Trigger systems are implemented at multiple levels, with ultra-fast but simple algorithms in hardware (L1) and increasingly complex algorithms in software at higher levels (HLT), where a more detailed readout of the detectors is exploited.
An event is fully recorded only if it satisfies the selection criteria at all levels of the trigger.
The L1 triggers decrease the 40 MHz rate down to $\mathcal{O}(100)$ kHz,
which is further reduced to $\mathcal{O}(1)$ kHz after the HLT triggers.
In order to achieve these reductions, triggers targeting processes with a very high cross-section are prescaled:
events are randomly discarded so that only a fraction $1/p$ ($p=\text{prescale}$) are recorded.
For example, at $\sqrt{s}=8$ TeV, the ATLAS single jet triggers targeting events with at least one jet with a minimum transverse momentum 
($p_\text{T}$) between $50$ and $100$ GeV had prescales of $p\sim 10^4$~\cite{Aad:2015cua}.
The lowest unprescaled ($p=1$) single jet trigger requires a minimum transverse momentum $p_\text{T}\sim500$~GeV.
The prescales for low $p_\text{T}$ processes, such as inclusive jet production, 
are increased at least linearly with the instantaneous luminosity, in order to keep the rate constant.  

While the existing trigger system is very effective at identifying high-$p_\text{T}$ objects, there are a plethora of viable models of physics beyond the Standard Model (BSM) that are not well covered.  One broad class of models predicts exotic signatures involving isolated charged particle tracks.  Pattern recognition for track reconstruction in the ATLAS and CMS inner detectors is computationally expensive and it only runs in a limited way at the HLT (and possibly at L1 in the future).  Tracks with kinks, displaced vertices, high dE/dx, anomalous timing, intermittent hits, and exotic curvature will not be efficiently reconstructed by L1 tracking and are also not easily (or at all) covered in the HLT (see Ref.~\cite{Meade:2011du} for a review).  For example, oscillating pairs of tracks from new strong dynamics~\cite{Okun:1980kw,Okun:1980mu,Kang:2008ea} require dedicated reconstruction algorithms.  In addition to models with low multiplicity tracks, BSM processes that predict extreme multiplicities of low energy particles~\cite{Knapen:2016hky,Harnik:2008ax,Kang:2008ea} are striking signatures that may be largely uncovered by existing or even possible triggering techniques.  Another broad class of models predicts exotic structure inside hadronic jets.  This includes jets with many displaced vertices~\cite{Schwaller:2015gea} as well as jets with large invisible components~\cite{Cohen:2015toa}.  There are likely many other models that have yet to be proposed in the literature that would leave extraordinary detector signatures but too exotic to be captured by standard trigger schemes.  

All of the models discussed so far have the property that their signature is so exotic that there is likely not a significant background rate from the Standard Model.  The current triggering scheme also limits the sensitivity to models with large SM background, such that large prescales are required.  This includes low mass dijet resonances, such as a leptophobic $Z'$~\cite{Dobrescu:2013coa}.

There are two existing strategies for recovering model coverage that would otherwise have been lost by the trigger.  One strategy is to look for a target process produced in association with another 
very energetic object that can be used for triggering.
For example, a low mass $Z'$ that decays into jets can be produced in association with a high $p_\text{T}$ 
photon~\cite{ATLAS:2016jcu} or jet~\cite{cmsdijet} from initial state radiation (ISR).
However, this strategy introduces a large effective prescale due to a reduction in the cross-section.  
In addition, this procedure cannot be used to measure new or Standard Model processes differentially in their low $p_\text{T}$ phase space.  
Another powerful strategy, referred to as data-scouting or trigger-level analysis, stores only a smaller relevant fraction of the detector information for the events 
selected by the L1 trigger~\cite{Khachatryan:2016ecr,ATLAS:2016xiv}.
These trigger-level analyses are not impacted by the prescales of the HLT triggers, 
but are limited by the L1 prescales that are often tighter.
As only a small fraction of the detector information can be recorded at the L1 accept rate, only the specific final states for which the trigger-level analysis strategy has been designed are accessible.

The new strategy presented in this paper uses each individual pileup interaction for physics analysis.
All of these interactions potentially contain interesting physics processes and are recorded by the detector along with the primary interaction that satisfied any arbitrary trigger. Every event passing any trigger can be used for the purpose of studying the pileup interactions, which are recorded with nearly no selection bias. The effective prescale associated to this Zero Bias Sample (ZBS) is inversely proportional to the overall trigger bandwidth.  For a sufficiently high bandwidth, this effective prescale can be lower than the one from the ISR strategy.  In addition, a trigger-level analysis can be combined with this strategy, thus enhancing even further the physics reach by enabling access to a large quantity of otherwise unused data.  For analysis offline, complex reconstruction algorithms can run without real-time constraints on algorithm speed that would have been required to save the event using a targeted trigger online.

\section{The Zero Bias Sample}
\label{sec:zerobias}

Reconstructed tracks from charged particles are the most important handle for identifying pileup interactions.
Individual collision vertices are built from tracks and various objects can be associated to these vertices 
through their associated tracks.
For example, the jet vertex tagger (JVT) used in ATLAS is 90\% efficient at associating a jet with $20 <p_\text{T}< 50$ GeV 
to its correct vertex while mis-identifying stochastic or QCD jets from other vertices 1\% of the time~\cite{Aad:2015ina}.
Ignoring the small detector inefficiencies and fake rates, 
the effective luminosity from the pileup collisions collected from a trigger system with bandwidth $w$ is given by

\vspace{5mm}

\begin{align}
\label{eq:ZBSequation}
\int \mathcal{L}\text{(ZBS)} dt = \frac{w}{\text{40 MHz}}\times \int \mathcal{L}dt.
\end{align}

\vspace{5mm}

\noindent The last term in Eq.~\ref{eq:ZBSequation} 
is the integrated luminosity collected with standard triggers. 
One way to derive this equation is to consider the number of events recorded by the LHC experiments.  
Suppose there is a process $X$ with a cross-section $\sigma_X$.  
If $\Sigma$ is the total inelastic cross-section, then a fraction $\sigma_X/\Sigma$ (on average) of all $pp$ collisions recorded will contain the process $X$.  
If the bandwidth is $w$ and assuming that the production of $X$ does not significantly influence this rate, 
then the total number of $X$ events will be $\langle\mu\rangle\times w\times T\times \sigma_X/\Sigma$, 
where $T$ is the amount of time the LHC is operated for $pp$ collisions and $\langle \mu\rangle$ is the average number of $pp$ collisions per bunch crossing.
Now let $Y$ be any Standard Model process that can be triggered with 100\% efficiency and has a prescale of 1.
Analogously to the calculation for $X$, the number of $Y$ events recorded will be $\langle\mu\rangle\times H\times T\times \sigma_Y/\Sigma$, 
where $H$ is the total rate of bunch crossings (40 MHz at the LHC).  
This shows that $\int \mathcal{L}dt=\langle\mu\rangle\times H\times T / \Sigma$ and 
solving the system of equations to derive the effective luminosity for the process $X$ results in Eq.~\ref{eq:ZBSequation}. 
The bandwidth $w$ can go from the 100-500 Hz typical of Run 1 (2010-2012) data-taking to the upper limit of the expected
L1 trigger bandwidth for the High Luminosity LHC (HL-LHC) experiments~\cite{CERN-LHCC-2015-020,Butler:2055167}.
Table~\ref{tab} shows the ZBS integrated luminosity for various LHC running conditions in Run 1 and projected for Run 2 (2015-2018) and beyond.
The effective prescale for the best HL-LHC data acquisition scenarios (last row in Table~\ref{tab}) is between 4000 and 6000.
For trigger-level analyses of the ZBS dataset at the HL-LHC, the effective prescale is between 50 and 100.  This means that if a particular signature has a primary trigger efficiency that is less than $0.02$-$0.03$\% offline or $1$-$2\%$ at L1, then the ZBS dataset will record more signal events.  Note that the predicted integrated luminosity during Runs 2 and 3 of the LHC is about 300 fb$^{-1}$ while the final HL-LHC data is expected to be about 3 ab$^{-1}$.   


\begin{table}[h!]
\begin{center}
\begin{tabular}{| m{15mm} | m{8mm} | m{6mm} | m{10mm} | m{10mm} | m{16mm} | m{10mm} |}
\hline
\centering LHC Run & \centering Total Lumi. [1/fb]  &\centering $\langle \mu\rangle$ &\centering L1 Rate [MHz] &\centering HLT Rate [kHz]&\centering ZBS [1/fb] &\centering ZBS @ HLT [1/fb]\tabularnewline
 \hline
 \hline
\centering 1 &\centering 20 &\centering 20 &\centering 0.1  &\centering 0.1  &\centering $5\times 10^{-5}$ &\centering 0.05 \tabularnewline
\centering 2+3 &\centering 300 &\centering 80 &\centering 0.1  &\centering 1  & \centering$7.5\times10^{-3}$ &\centering 0.75 \tabularnewline
\centering 4+5 (ATLAS) &\centering 3000 &\centering 200 &\centering 0.4  &\centering 10  &\centering 0.75 &\centering 30 \tabularnewline
\centering 4+5 (CMS) &\centering 3000 &\centering 200 &\centering 0.75  &\centering 7.5  &\centering 0.56 &\centering 56.3 \tabularnewline
 \hline
\end{tabular}
\end{center}
\caption{The ZBS integrated luminosity for various LHC running conditions in Run 1 and projected for Run 2 and beyond.
The last column shows the projected luminosity when performing a trigger-level analysis, before the application of any software prescales.
}
\label{tab}
\end{table}

\section{Cross-section Projections}
\label{sec:lowmass}

\begin{figure}[h!]
\centering
\includegraphics[width=0.5\textwidth]{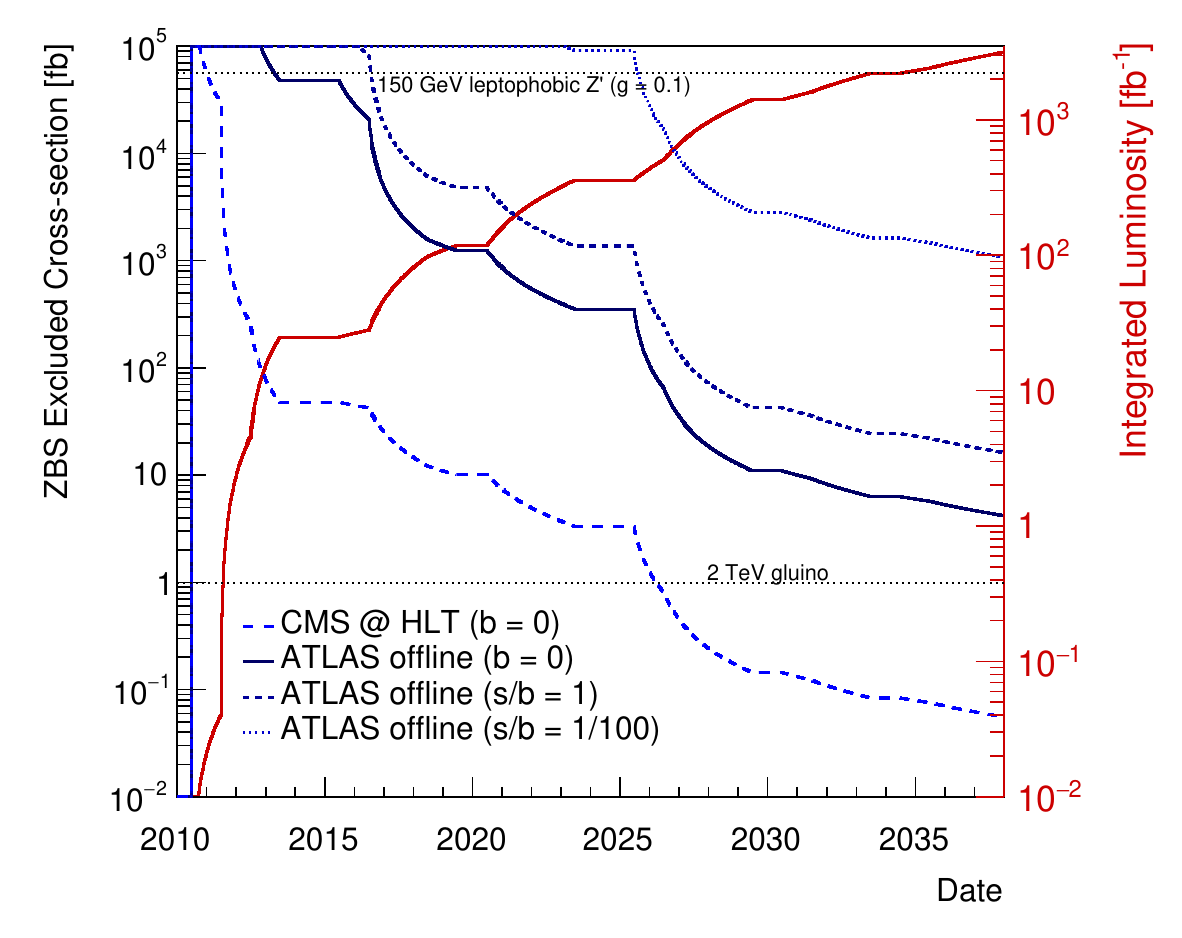}
\caption{The excluded cross-section and the integrated luminosity are shown throughout the lifetime of the (HL-)LHC.  Statistical uncertainties are assumed to dominate over systematic ones in this low background regime.  For reference, the cross-section for a 2 TeV gluino and a 150 GeV leptophobic $Z'$ with $g=0.1$ are also shown (see text for details).}
\label{fig:power}
\end{figure}

Figure~\ref{fig:power} presents the 95\% confidence level cross-section limits with the $\text{CL}_\text{s}$ procedure~\cite{Read:2002hq}, covering a broad class of models, for the ZBS throughout the lifetime of the (HL-)LHC.  Limits are computed assuming various signal-to-background ratios.  For a zero-background search, at least three signal events are needed to reach the $95\%$ confidence limit.  

The offline version of the ZBS can be analyzed at any time, including (well) after the full HL-LHC program has ended, while the trigger-level (online) version requires some level of analysis to be implemented in the software trigger at run time.   For a model that predicts an extraordinary signature with nearly no SM background (see Sec.~\ref{sec:intro}), a ZBS analysis will be able to set a cross-section limit of nearly 3 fb.  For reference, this is the cross-section of a 1.8 TeV gluino~\cite{Borschensky:2014cia}, a 1.1 TeV stop~\cite{Borschensky:2014cia}, a 1.4 TeV ferminonic top quark partner~\cite{Czakon:2011xx}, and a 1.7 TeV colored quirk with infracolor representation size 5~\cite{Kang:2008ea}.  Cross-section limits for the ZBS applied at HLT are nearly a factor of 100 stronger, though would require dedicated algorithms to be put in place before the start of the HL-LHC to achieve the full potential.   Figure~\ref{fig:power} also shows the limits for searches that have a non-trivial background component.  While many of the models described in the introduction are nearly background-free, some may have contributions from low probability events from the Standard Model.  With a signal-to-background ratio of $0.001$, a cross section limit of about 1 pb can be set.

\begin{figure}[h!]
\centering
\includegraphics[width=0.45\textwidth]{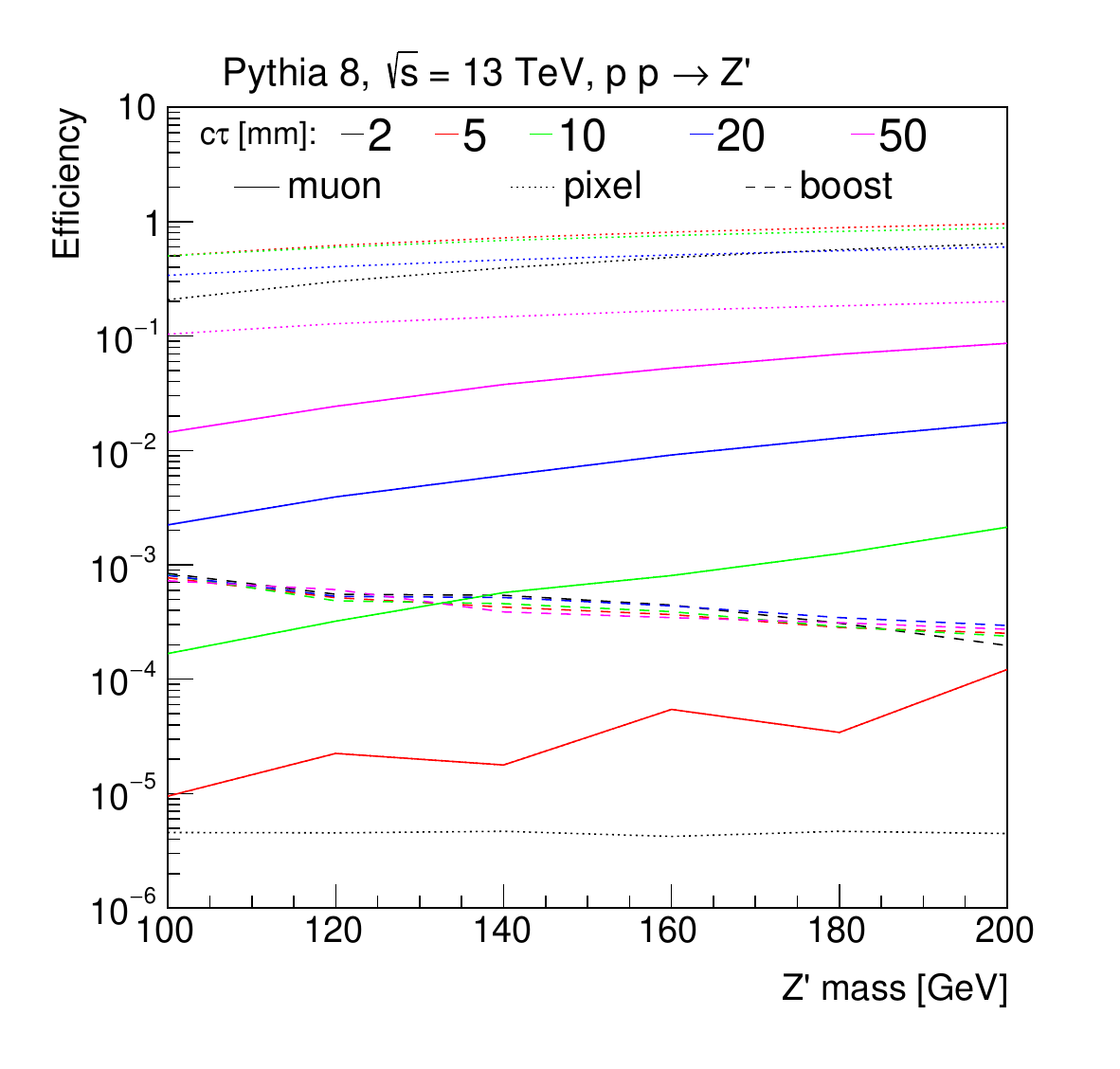}
\caption{The efficiency for three analysis strategies for identifying $Z'$ decays into dark quarks that subsequently decay into SM mesons.  The dotted line for the 2 mm muons indicates an upper bound based on nearly one million events per mass point.  See the text for further details.}
\label{fig:hiddenvalley}
\end{figure}

The important low mass dijet resonance search is a concrete illustration of the power of the ZBS.   Every hadron collider has searches for dijet resonances, which are predicted in a wide variety of BSM models.  To start, consider a case with a high signal-to-background ratio.  For example, suppose the dark matter consists of an extended sector of quark- and gluon-like objects and a confining QCD-like SU(3)-symmetry as in e.g.~\cite{Schwaller:2015gea}.  In certain regions of the model parameter space, such a model would give rise to emerging jets wherein jets are formed in the dark sector and then after some time, the dark quark and gluon fragmentation products decay into SM particles.  Suppose there is a leptophobic Z' that connects the visible sector QCD with the dark sector QCD.  As a minimal but complex realization of this model~\footnote{Pythia 8.2~\cite{Sjostrand:2014zea} includes a Hidden Valley~\cite{Strassler:2006im} module that even incorporates the running of the dark $\alpha_s$~\cite{Carloni:2010tw,Carloni:2011kk}.  However, when the dark quarks (\texttt{qv}) are very light (SM quark-like) then the decay and hadronization can be eliminated.  As an alternative, two instances of Pythia 8 are used to generate $Z'$ events, one with $Z'\rightarrow qq'$ and one with $Z'\rightarrow \nu\bar{\nu}$.  The first instance has modified SM parameters.  The hadrons resulting from the modified SM case (possible since the $Z'$ is a color-singlet) are identified and then placed in an unmodified event but with the vector sum boosted to match the $Z'\rightarrow \nu\bar{\nu}$ four-vector.  In the modified setup, there are no baryons, quarks do not radiate photons, electromagnetically decaying hadrons are stable, and the dark pions have the same mass as the SM $\rho$ meson so that they can decay into to two SM pions.  The modified parameters are \texttt{TimeShower:QEDshowerByQ  = false}, \texttt{StringFlav:decupletSup = 0}, \texttt{StringFlav:probStoUD = 0}, \texttt{StringFlav:probQQtoQ = 0}, \texttt{StringFlav:etaSup = 0}, \texttt{StringFlav:etaPrimeSup = 0}, \texttt{211:m0:0.775}, \texttt{111:m0:0.775}, \texttt{111:mayDecay = false}, \texttt{223:mayDecay = false}, \texttt{221:mayDecay = false}.}, the dark sector is a nearly exact copy of the SM QCD where, for simplicity, there is only one hadron called the dark $\rho$.  There are two relevant free parameters of this model for studying various strategies: the mass of the $Z'$ and the lifetime of the dark hadrons $c\tau$, which is set by the couplings of the $Z'$.   Each dark $\rho$ resulting from dark quark or gluon fragmentation could result in a displaced vertex.  Figure~\ref{fig:hiddenvalley} shows the efficiency of various identification algorithms as a function of $Z'$ mass for various lifetimes.  The following paragraph explains and compares the three schemes, where the first (`pixel') scheme uses the ZBS while the other two use traditional triggering strategies. 

The first possibility is to reconstruct displaced vertices in the pixel systems of the ATLAS or CMS detectors, labeled~\footnote{For this case, the decay is required to occur between 30 and 100 mm.  No material veto is applied, which in practice will lower the efficiency.} `pixel' in Fig.~\ref{fig:hiddenvalley}.   It is not possible to accurately estimate the background from simulation, but based on Refs.~\cite{Aad:2015rba,CMS:2014wda,Aaboud:2017iio}, it seems conservative that requiring $\geq 4$ displaced vertices is near the zero-background regime.  The maximum efficiency is for $c\tau\sim  5$ mm and is $\gtrsim 50\%$ across the entire mass range.  The ZBS may be a powerful tool to target these models because standard approaches are not powerful: there is usually not enough $E_\text{T}^\text{miss}$ or $H_\text{T}$ to trigger on, as in the SUSY cases studied in Refs.~\cite{Aad:2015rba,CMS:2014wda,Aaboud:2017iio}.  Instead, if the dark mesons decay in the muon chambers of ATLAS or CMS, then the muon trigger could be used to identify events~\cite{Aad:2015uaa,Aad:2013txa}, indicated by~\footnote{For concreteness, this case corresponds to any decay that happens within the ATLAS muon spectrometer (MS) acceptance, between 4 and 7 meters (similar results hold for CMS, where the MS is closer to the interaction point).  Pions from the dark $\rho$ decay must be at least 10 GeV~\cite{Aad:2013txa}.  The actual trigger efficiency will be much lower, because the particles.} `muon' in Fig.~\ref{fig:hiddenvalley}.  A third strategy is to identify cases in which the $Z'$ is sufficiently boosted that all of its decay products are captured inside a single jet.  In that case, one could look for a bump in the single jet mass distribution~\cite{cmsdijet}, indicated by~\footnote{In particular, this case is defined by $2m^{Z'}/p_{T}^{Z'}\leq 0.8$.  Further selection criteria on the number of prongs in the jet will further reduce the efficiency and even with strict criteria, the search does not approach zero-background because there is a large background from `normal' QCD jets.  The power of this strategy is further reduced for $Z'$ with large width (in Fig.~\ref{fig:hiddenvalley}, the width is assumed to be negligible).} `boosted' in Fig.~\ref{fig:hiddenvalley}.  There is a large effective prescale from requiring the $Z'$ to be boosted.

The prescale for the offline ZBS for ATLAS is 4000 (Table~\ref{tab}); therefore, the efficiency for the ZBS and the displaced vertex approach is about $50\%/(4000)=10^{-4}$ for $m_{Z'}=200$ GeV and $c\tau\sim 5 $ mm.  The muon trigger has a similar or slightly lower efficiency when $c\tau$ is $\lesssim 5$ mm.  Both projections have some approximations and to know precisely which is better would require a detailed detector and background simulation.  These studies indicate that both methods would result in similar sensitivity and therefore the ZBS strategy is worth pursuing further.  This is especially true if one can implement region of interest secondary vertex reconstruction at HLT.  The boosted strategy is likely too inefficient, as the necessary condition of the opening angle for the decay is already at the $10^{-4}$ level at $m_{Z'}=200$ GeV.

\begin{figure}[h!]
\centering
\includegraphics[width=0.45\textwidth]{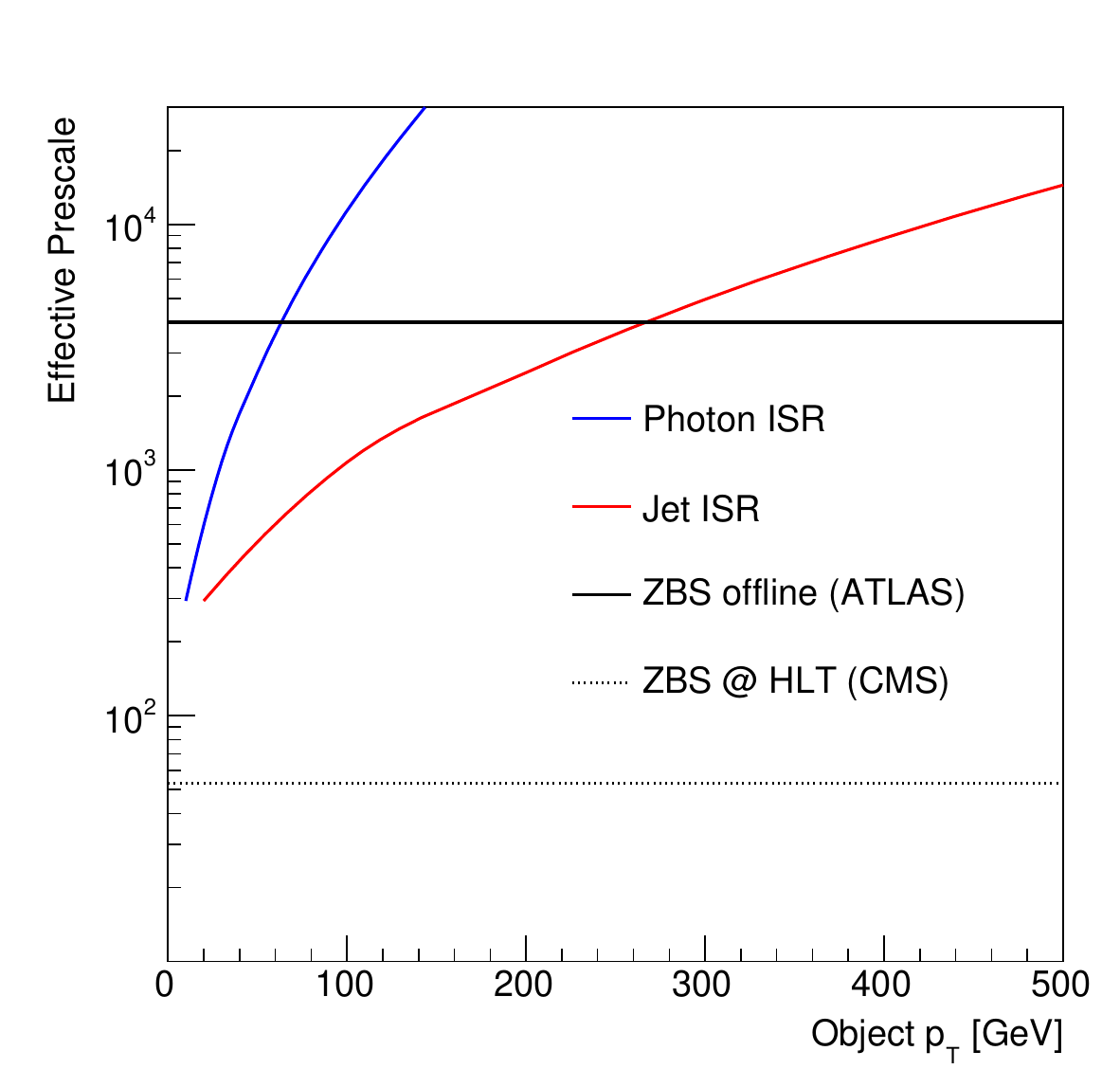}
\caption{A comparison of the effective prescale for the ZBS and ISR analyses for a particular leptophobic Z' model~\cite{Dobrescu:2013coa}.  The ZBS prescales are the same as in Table~\ref{tab}. }
\label{fig:prescale}
\end{figure}

In addition to low background searches for exotic signatures, the ZBS can also be competitive with searches for SM-like final states that exploit associated production. To illustrate this case, consider the traditional low-mass dijet resonance search.  The SM dijet cross-section is so large that searching for bumps in the dijet invariant mass ($m_{jj}$) spectrum is plagued by large prescales at low $m_{jj}$. Such searches are therefore performed with trigger level and ISR analyses.
Figure~\ref{fig:prescale} shows the effective ZBS prescale\footnote{Prescales for the ISR analyses are computed with MG5\_aMC 2.1.1~\cite{Alwall:2014hca} by comparing the cross-section for \vtext~or \vtexta~to \vtextb and do not include any further prescale obtained at L1 or HLT.} compared with the effective prescale due to the reduction in
cross section when requiring the $Z'$ to be produced in association with a ISR high $p_\text{T}$ photon or jet.  By construction, the prescale is independent of $p_\text{T}$ for the ZBS, but grows quickly with the photon or jet $p_\text{T}$ for the ISR analyses. Typical minimum
requirements for unprescaled single photon and jet triggers are $p_\text{T}=100$ and $p_\text{T}=400$ GeV respectively. At these values
the effective prescale is significantly larger than the one expected for the ZBS.  The photon and jet thresholds can be lowered when combining the ISR technique with data scouting.  However, when the ZBS is combined with data scouting (at the HL-LHC), the solid line in Fig.~\ref{fig:prescale} becomes the dashed one.  Therefore, even at trigger level, the ZBS analysis has a lower effective prescale than ISR techniques.

The ZBS is superior to the ISR technique in terms of prescale (at the HL-LHC), but a fair comparison also requires an assessment of the signal-to-background ratio.  A loss in events from the effective prescale from an ISR requirement can be partially compensated by better discrimination power.  To estimate the approximate sensitivity to a dijet resonance, a benchmark $Z'$ model~\cite{Dobrescu:2013coa} 
is simulated with MG5\_aMC 2.1.1~\cite{Alwall:2014hca} interfaced with Pythia 8.170~\cite{Pythia8}.
To simulate the detector response, the jet momenta are smeared according to 
$\sigma(p_\text{T})/p_\text{T} = 1.3/\sqrt{p_\text{T}/\text{GeV}}$.
The jet resolution depends on the pileup conditions and is in general worse at trigger level 
than for fully reconstructed offline jets.
Therefore, the resolution function is conservatively chosen to be worse than 
the typical energy resolution at LHC experiments in Run 2.
Events are required to have two jets with $p_\text{T}>25$ GeV, and the two leading such jets are used to 
compute the dijet invariant mass, $m_{jj}$.
More sophisticated approaches could better exploit events with significant initial or final state radiation
for an enhanced sensitivity but are beyond the scope of this paper.
A simple binned $\chi^2$ analysis of the dijet invariant mass spectrum in a window around the target $Z'$ mass is performed, 
using toy Monte Carlo simulation to estimate the $p$-value.
A given mass point is declared excluded if the corresponding $p$-value is less than $0.05$.
As a validation of this procedure, the coupling upper limit is estimated for a $500$ GeV $Z'$ with $20.3$ fb${}^{-1}$
of unprescaled single jet trigger simulated data at $\sqrt{s}=8$ TeV.
The limit obtained, approximately $1.5$, is consistent with the Run 1 ATLAS result~\cite{Aad:2014aqa}.

\begin{figure}[h!]
\centering
\includegraphics[width=0.45\textwidth]{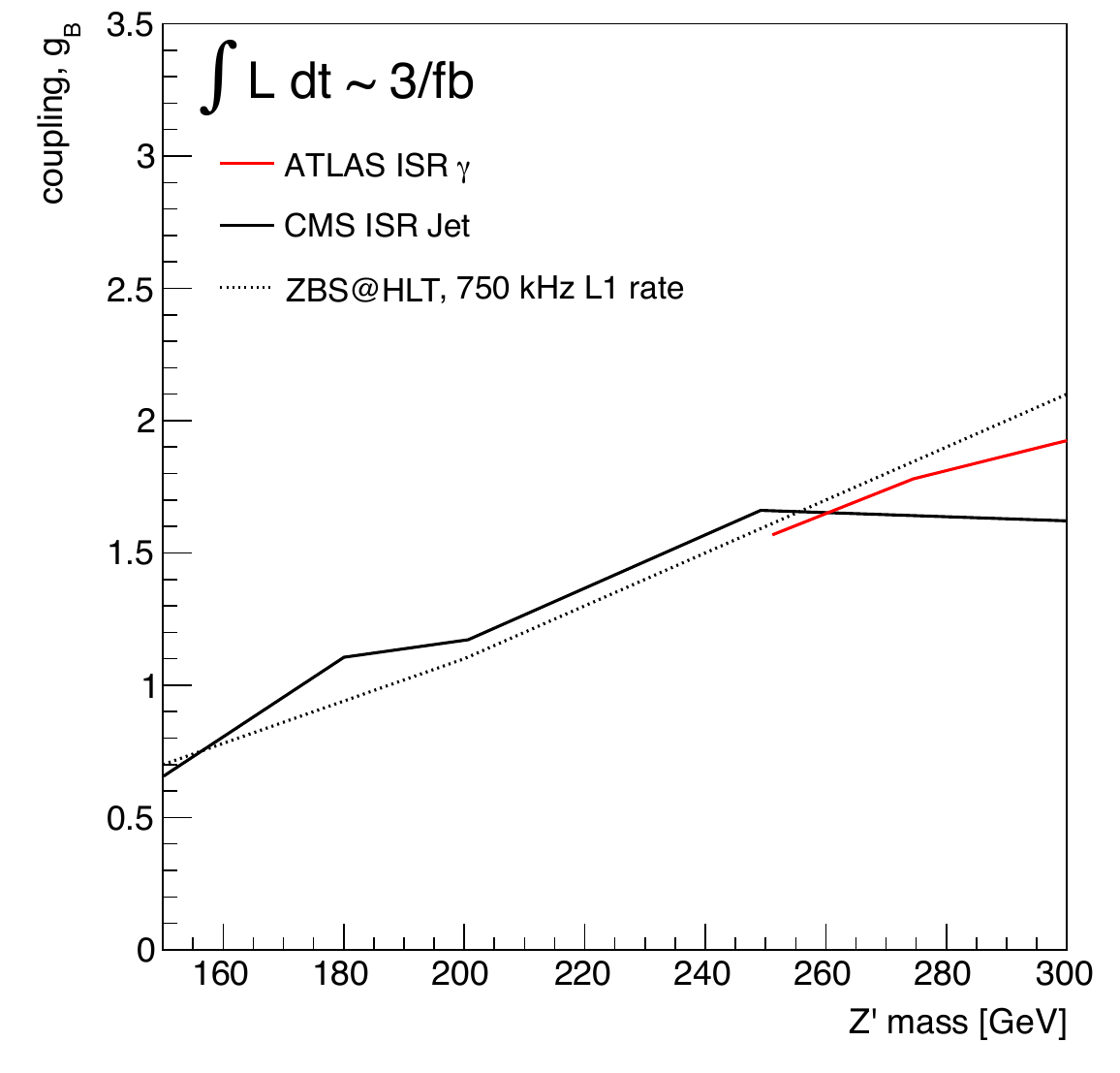}
\caption{The coupling limits as a function of the $Z'$ mass for ISR jet/photon analyses using the 2015 LHC run datasets compared with estimates for the ZBS using a dataset of the same size as collected at the HL-LHC (with a significantly higher bandwidth).  Existing limits are from Ref.~\cite{cmsdijet,ATLAS:2016jcu}.}
\label{blah2}
\end{figure}

Figure~\ref{blah2} shows a comparison between published ISR limits and our estimate of the ZBS @ HLT based on the standard search for a peak in the $m_{jj}$ distribution.  The ISR result will slowly improve with more integrated luminosity so for a fair comparison, both strategies are evaluated with a dataset size corresponding to the 2015 run.  The ZBS results are estimated assuming that the same dataset is recorded at HL-LHC rates so a relative comparison between strategies near their peak performance is possible. With this setup, the limits are found to be comparable.  Given the complementarity of the two strategies, further gain can be achieved by combining the results.  Note that a conservative estimate of the jet resolution at the HL-LHC is used for the ZBS analysis (which is also highly simplified).  It is likely that the limits shown here are therefore conservative for the ZBS (also supported by Fig.~\ref{fig:prescale}). 


\section{Implementation Challenges}

While the ZBS dataset holds great promise, using the data in practice will be technically challenging.   The first challenge is to access data from all triggers.  This is both a challenge for a ZBS analysis online and offline.  Online, the problem is that most HLT items are tied to a single L1 trigger.   ZBS analyses working at the trigger level would need to have access to all events that pass L1, which is a bandwidth challenge.  Offline, the problem is that the data are separated into streams and most analyses require a single trigger in order to reduce the data volume.  Overcoming these challenges would require more sophisticated bookkeeping algorithms.  

The second challenge is having access to all of the pileup information.  The time to perform track reconstruction does not scale well with $\mu$ and so current fast algorithms operating at the trigger level reduce the bandwidth by explicitly ignoring pileup and displaced tracks as early as possible in the analysis chain (see e.g. Ref.~\cite{Shochet:1552953}).  Special and possibly time consuming track reconstruction may be required for some analyses that are looking for exotic track signatures in the ZBS.  This is a more serious challenge online, where information about the pileup collisions may even be discarded before it can ever be analyzed.  For example, the current trigger-level analysis procedure (to reduce the data rate) is to save only what is needed for the final analysis selection for offline processing.   If after the data are collected, there is an idea that can be studied with the ZBS, it will not be possible to look at other information in these data.  Therefore, analyses that use the ZBS at HLT will need to be designed prior to data taking.  Even if they are designed ahead of time, accessing and reconstructing all of the pileup information at the HLT will be a significant technical challenge.  Both ATLAS and CMS have ambitious goals for fast event reconstruction and it seems possible, though with much effort, to make the ZBS analysis at HLT work.

Another technical challenge is that most algorithms for event reconstruction are designed for a single primary vertex; this would need to be generalized to handle any vertex.  For example, the fraction of track energy from the hard scatter collision that is used to identify pileup jets needs to be recomputable with respect to any vertex labeled as hard scatter.  Some of these tools already exist, such as the re-assignment of the hard-scatter vertex in the $H\rightarrow\gamma\gamma$ measurement in ATLAS that can use calorimeter pointing information to identify the correct vertex~\cite{Aad:2012tfa}.  The reconstruction resolution will always be a challenge at high $\mu$, but a jet with a given $p_\text{T}$ from a pileup collision will have the same resolution as a jet with the same $p_\text{T}$ from the collision that triggered the event.  Therefore, this is a challenge that the primary trigger analyses will also face.

None of these challenges exclude the possibility of a ZBS analysis, but they do show that while the data are will be produced `for free', significant effort will be required to ensure they are collectable and analyzable. 

\section{Conclusions}
\label{sec:conclusions}

The multiple pileup interactions produced in LHC collisions yield unbiased data which can 
be used to probe physics processes otherwise unaccessible or with limited acceptance.  The effective prescale for this Zero Bias Dataset is about 40000 in Runs 2+3, and drops to 400 for trigger-level analyses.  If the trigger efficiency for any search is lower than this amount, then the ZBS may be more powerful.  In particular, for exotic signatures that are nearly impossible to trigger on due to bandwidth and time constraints in the trigger, the ZBS may be the best strategy.  This was illustrated explicitly for a Hidden Valley model with a $Z'$ and dark sector QCD, where the ZBS dataset has a sensitivity that is likely comparable to or better than existing trigger strategies.  Of course, the ZBS idea will apply to models that have not yet appeared in the literature.  For existing models, one can fully exploit the ZBS by implementing selections in the software trigger to set the most stringent limits despite failing a direct hardware trigger.  As discussed in the previous section, using the ZBS data would be technically challenging for both the online and offline versions.  The studies and examples presented in earlier sections show that these costs in time and effort are worth serious consideration.

In addition to setting a bottom line for searches with a low background rate, the ZBS may also be competitive with traditional searches that exploit associative production to pass the trigger.  Requiring a second object, such as a high-$p_\text{T}$ ISR jet or photon introduces a large effective prescale that can be harsher than the prescale from the ZBS.  When combined with a trigger-level analysis, the ZBS is expected to provide comparable limits to the ISR technique in the case of the low mass dijet resonance search. These estimates are based on a simplified model of the dijet resonance searches and could be improved with additional sophistication.  The simple model ignores the efficiency for reconstructing primary vertices and any inefficiencies in associating jets to these vertices.  For the relatively high masses targeted by the example, these inefficiencies are relatively small.  However, for lower mass measurements and searches, these inefficiencies may be important.  Both ATLAS and CMS have ongoing studies to improve the tracking performance in high pileup environments, including the use of timing information to distinguish objects from spatially overlapping vertices.  All of these interesting developments will be important for the ZBS strategy.

After the full LHC program, the ZBS will have accumulated about $0.5$-$1$ fb${}^{-1}$ of fully unbiased $pp$ collision data that would not have been analyzed.  We have shown that the novel concept of analyzing all pileup interactions enhances the physics reach of the LHC experiments and could constitute a useful strategy to fully exploit the HL-LHC dataset.

 
\section{Acknowledgments}

We would like to thank Felix Yu for help simulating the $Z'$ benchmark model and Nhan Tran for useful discussions about the ZBS.  In addition, we are grateful to John Alison for helping us compute the correct luminosity for the ZBS.

\bibliography{myrefs}

\end{document}